\documentclass[usenatbib]{mn2e}
\usepackage{graphicx}

\title[Site Testing]{Sky Quality Meter and satellite correlation for the night cloud cover analysis at astronomical sites}

\author[S. Cavazzani et al.]{S. Cavazzani$^{1,2}$\thanks{E-mail:stefano.cavazzani@unipd.it}, S. Ortolani$^{1,2}$, A. Bertolo$^{3}$, R. Binotto$^{3}$, P. Fiorentin$^{4}$, 
\newauthor G. Carraro$^{1}$, I. Saviane$^{5}$ and V. Zitelli$^{6}$
\\
$^{1}$Department of Physics and Astronomy, University of Padova, Vicolo
dell'Osservatorio 3, 35122, Padova, Italy\\
$^{2}$INAF-Osservatorio Astronomico di Padova, Vicolo dell'Osservatorio 5, 35122, Padova, Italy\\
$^{3}$ Regional Environmental Prevention and Protection Agency of Veneto, Via Ospedale Civile 24, 35121, Padova, Italy\\
$^{4}$Department of Industrial Engineering, University of Padova, via Gradenigo 6a, 35131, Padova, Italy\\
$^{5}$European Southern Observatory, Alonso de Cordova, 3107, Santiago, Chile\\
$^{6}$INAF-OAS Osservatorio di Astrofisica e Scienza dello Spazio di Bologna, Via Gobetti 93/3, 40129, Bologna, Italy
}

\begin{document}

\date{Accepted 0000 September 00.  Received 0000 September 00; in original form 0000 May 00.}

\pagerange{\pageref{firstpage}--\pageref{lastpage}} \pubyear{2009}

\maketitle

\label{firstpage}

\begin{abstract}

The analysis of the night cloud cover is very important for astronomical observation in real time, considering a typical observation time of about 15 minutes, and to have a statistics of the night cloud cover.
In this paper we use the SQM (Sky Quality Meter) for high resolution temporal analysis of the La Silla and Asiago (Ekar observatory) sky: 3 and 5 minutes respectively. We investigate the annual temporal evolution of the natural contributions of the sky in a site not influenced by artificial light at night (ALAN) and one highly influenced respectively. We also make a correlation between GOES and AQUA satellites data and ground-based SQM data to confirm a relationship between the SQM data and cloud cover. We develop an algorithm that allows the use of the SQM for night cloud detection and we reach a correlation of 97.2\% at La Silla and 94.6\% at Asiago with the nighttime cloud cover detected by the GOES and AQUA satellites. Our algorithm also classifies the photometric (PN) and spectroscopic nights (SN). We measure 59.1\% PN and 21.7\% SN for a total percentage of clear nights of 80.8\% at La Silla in 2018. The respective Ekar observatory values are 31.1\% PN, 24.0\% SN and 55.1\% of total clear nights time. Application to the SQM network would involve the development of long-term statistics and big data forecasting models, for site testing and real-time astronomical observation. 

\end{abstract}


\begin{figure*}
  \centering
  \includegraphics[width=16cm]{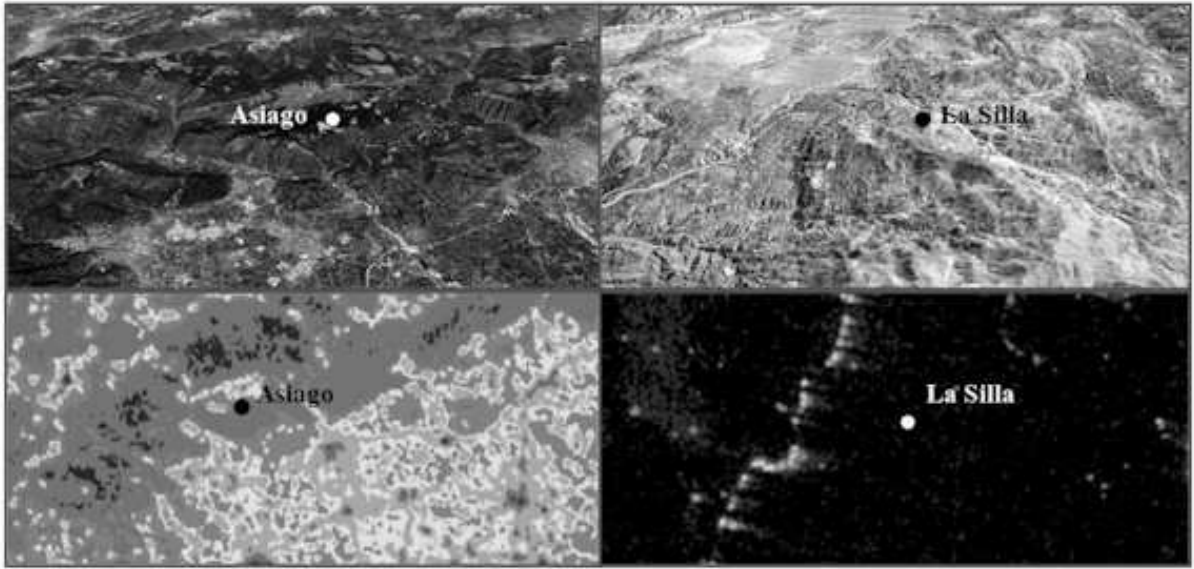}
  \caption{Location of the analysed site. The top left panel shows the topographical characteristics of Asiago, the point indicates the Ekar observatory, the bottom left panel shows the average light emission in 2018 for the Asiago area detected by VIIRS. The panels on the right show the same characteristics for La Silla observatory.}
             \label{map}
\end{figure*}

\section{Introduction}

Artificial light at night (ALAN) increases the night sky brightness, creating the greatest visible effect of light pollution and in particular influencing the astronomical observations in contaminated sites.
In the last decades the light pollution has become a global scale phenomenon (Kyba et al., \cite{ky}) as evidenced by the growing interest from scientists in fields of astronomy (Patat et al., \cite{pat}; Puschnig et al., \cite{pu}; Zhang et al., \cite{zha}), ecology, biology (Holker et al., \cite{hol}; Gaston et al., \cite{gas}; Manfrin et al., \cite{man}) and medicine (Kloog et al., \cite{klo}; Stevens et al. \cite{ste}). 
The study of this strongly interdisciplinary subject has been rapidly increasing, as evidenced by the growing literature on the subject (Mulder et al., \cite{muld}).
Light pollution is produced by  two main components: the natural component, in turn divided into terrestrial and extraterrestrial,  and the artificial component caused by human activities (ALAN).
We analyse the Sky Quality Meter (SQM) data for two sites with a high temporal resolution at La Silla and Asiago (Ekar observatory) in 2018: 3 and 5 minutes respectively (see Table \ref{site}). The SQM is one of the main tools for the light pollution analysis. The main features are described in Cinzano (\cite{cinz05} and \cite{cinz07}).
Other tools are described in Hanel et al. (\cite{han}).
SQM networks are widely used, as described in Bertolo et al. (\cite{berto}), Espey et al. (\cite{esp}), Posch et al. (\cite{posh}) and Pun et al. (\cite{pun}).\\
Figure \ref{map} shows the topographical maps of the two sites with their respective night images (2018 average) captured by VIIRS (Visible Infrared Imaging Radiometer Suite) sensor aboard the joint NASA/NOAA Suomi National Polar-orbiting Partnership (Suomi NPP) satellite.
The choice of sites gives us the comparison between a site influenced by ALAN and one that is not contaminated.\\ 
We study the contribution of the main natural factors (e.g. Milky Way, Moon, zodiacal light, etc.) and also show how the cloud cover influences the readings of the SQM.
One of the pioneer of the clouds contribution study has been Roy Garstang (Garstang, \cite{gar}).\\
This study was further developed and linked to the new SQM networks in Bara' et al. (\cite{bara16} and \cite{bara19}), and in Ribas et al. (\cite{ribas}).
The contribution of clouds and their impact on the biosphere is also studied in Jechow et al. (\cite{jec}).\\
In this paper we develop an algorithm based on SQM data for the cloud cover analysis and we correlate this result with satellite data. We analyzed the VIIRS data to measure the mean magnitude in 2018 in clear sky conditions to empirically calibrate the model. We calculated the cloud cover at night through the GOES satellite (Geostationary Operational Environmental Satellite) and the AQUA satellite, in particular its MODIS (Moderate Resolution Imaging Spectro-radiometer) tool. The first is a geostationary satellite while the second is a polar satellite. Finally, correlated these results with the SQM data.\\
The application of our algorithm to the SQM network would collect long and short term statistics of sky brightness and cloud cover at night, two fundamental parameters for astronomical observation. We can extrapolate important information for real-time observation and forecast modeling using only the SQM tool.\\  
The layout of the paper is as follows: Section \ref{aq} illustrates the details of the SQM measurements and ancillary satellite data products; in Section \ref{b} we describe the used method to derive information about cloud cover from the SQM data; in Sections \ref{c} and \ref{d} we correlate our SQM-cloud detection algorithm with satellite data for the temporal analysis of the photometric and spectroscopic nights at La Silla and at Asiago respectively. In Section \ref{error} we associate the uncertainties with the correlations.
Finally, in Section \ref{e} we discuss the results and present our conclusions.

\begin{table}
 \centering
 \begin{minipage}{80mm}
  \caption{Geographic characteristics of the analyzed sites.}
   \label{site}
  \begin{tabular}{@{}lcccc@{}}
  \hline

  site      &LAT.&      LONG. & Altitude  \\
            &    &            &  km      \\
 \hline
 Asiago (Ekar observatory)     &   $45^{\circ}50'$  &  $11^{\circ}34'$  &  $1.366$     \\
 La Silla   &   $-29^{\circ}15'$  &  $-70^{\circ}43'$  &   $2.347$       \\
 \hline

\end{tabular}
\end{minipage}
\end{table}

\begin{figure}
  \centering
  \includegraphics[width=8cm]{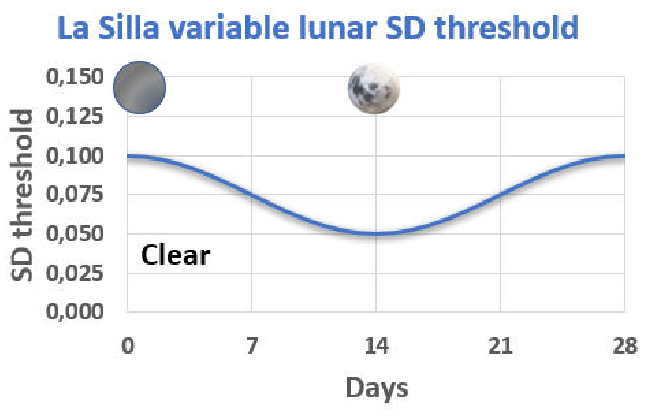}
  \includegraphics[width=8cm]{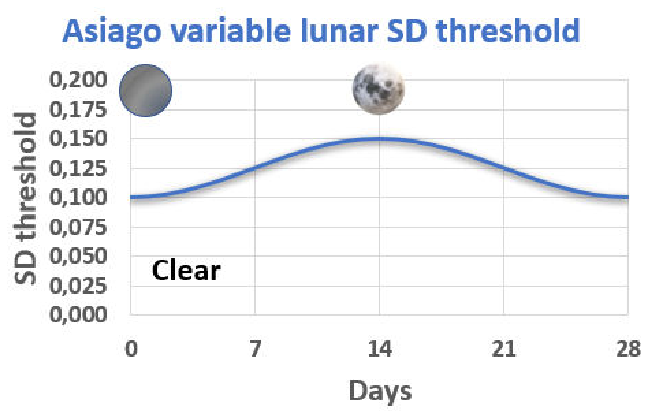}
  \caption{The top panel shows the standard deviation (SD) threshold trend according to the lunar cycles for a site not contaminated by ALAN (La Silla) while the bottom panel for a contaminated site (Asiago). The SD threshold is expressed in $mpsas=\frac{mag}{arcsec^{2}}$ (y-axis). In this case we considered a 28-day moon cycle.}
             \label{ath}
\end{figure}

\section{SQM measurements and ancillary satellite data products}
\label{aq}

The sky brightness measurements were carried out in both sites with a Sky Quality Meter-Lens Ethernet (SQM-LE) pointed to the zenith.
The SQM-LE measures the darkness of the night sky to provide readings of magnitudes per square arc
second ($mpsas=\frac{mag}{arcsec^{2}}$) through an Ethernet connection.
A light sensor provides the microcontroller with a light level, while the
the temperature sensor compensates the readings for various operating temperatures.
In this analysis we use the ancillary satellite data provided by GOES, Aqua/MODIS and VIIRS.\\
The Aqua satellite's orbit has a perigee of 691 km, an apogee of
708 km. Aqua MODIS view the entire surface
of the Earth every 1 to 2 d, acquiring data in 36 spectral bands, or
groups of wavelengths. The cloud cover is analysed with the bands
from 20 to 36 (see Table \ref{modis}). MODIS band 31 corresponds to the
wavelength of GOES band 4 (see Table \ref{goes}).
GOES satellite has a geostationary orbit at an altitude of 35800 km.

\begin{table}
 \centering
 \begin{minipage}{80mm}
  \caption{MODIS bands. The spatial resolution of the bands is $1 km$. }
   \label{modis}
  \begin{tabular}{@{}lccc@{}}
  \hline

 Primary Use	& Band      &  Bandwidth [$\mu m$] \\

 \hline
 	
Surface/Cloud Temperature	\\
&20  &   3.660 - 3.840	\\
&21	&   3.929 - 3.989	\\
&22	 &   3.929 - 3.989	\\
&23	 &   4.020 - 4.080 \\
\hline	
Atmospheric Temperature	\\

&24	& 4.433 - 4.498	 \\
&25	 &  4.482 - 4.549	\\

\hline

Cirrus Clouds Water Vapor	\\

&26	  & 1.360 - 1.390	\\
&27	  &  6.535 - 6.895	\\
&28	  &  7.175 - 7.475	\\
\hline
Cloud Properties \\

&29	 & 8.400 - 8.700	\\
\hline
Ozone	\\
&30	 & 9.580 - 9.880	\\
\hline
Surface/Cloud Temperature	\\

&31 &	10.780 - 11.280	\\
&32	& 11.770 - 12.270	\\
\hline
Cloud Top Altitude	\\
&33	& 13.185 - 13.485	\\
&34	 & 13.485 - 13.785	\\
&35	 & 13.785 - 14.085	\\
&36	 & 14.085 - 14.385	\\
\hline

\end{tabular}
\end{minipage}
\end{table}

\begin{table}
 \centering
 \begin{minipage}{80mm}
  \caption{GOES bands and resolution.}
   \label{goes}
  \begin{tabular}{@{}lcccc@{}}
  \hline
                & Window  & Passband     & Resolution      \\
                &         & $[\mu m]$     &          [km]           \\
 \hline
 \textit{BAND1} &Visible  &    $0.55\div0.75$ &  $4$\\
 \textit{BAND2} &   Microwaves  &     $3.80\div4.00$ & $4$  \\
 \textit{BAND3} &$H_{2}O$   &   $6.50\div7.00$ &  $4$\\
	\textit{BAND4}   & $IR$   & $10.20\div11.20$ & $4$\\
	\textit{BAND6}&   $CO_{2}$  &    $13.30$ &  $8$\\
 \hline

\end{tabular}
\end{minipage}
\end{table}

We processed GOES data using \textit{McIDAS-V}, a free software package (for model details see Cavazzani et al. (\cite{cava11}) and Cavazzani et al. (\cite{cava15})). In this analysis we use a single image per night (2:45, Local Time). This makes the GOES data highly comparable with MODIS data.
The MODIS data are analysed through Giovanni Interactive Visualization and Analysis Website\footnote{https://giovanni.gsfc.nasa.gov}. This tool is designed for visualization
and analysis of the atmosphere daily global $1^{\circ}\times 1^{\circ}$ products. There is a single image per night.
Finally, the VIIRS data provide the mean annual magnitude in clear sky conditions to calibrate the threshold.
The VIIRS sensor is a Suomi NPP satellite tool. The imaging day/night band (DNB) provides global data at 742m spatial resolution and is a calibrated radiometer. The DNB visible bands have a broad
spectral range of $0.5-0.9\mu m$ centered at $0.7\mu m$ and they have ability to collect low-light imagery at night.

\section{Methods and data analysis}
\label{b}

First we used the annual mean light emission detected by VIIRS above the two analysed sites. We use the light emission value converted into magnitude provided by the site Light pollution map\footnote{https://www.lightpollutionmap.info}. The site refers to the World Atlas of the artificial night sky brightness (Falchi et al., \cite{fal}) and it gives a mean value in clear conditions of 21.9 for La Silla and 20.9 for Ekar observatory in 2018. We use the derived magnitude to empirically calibrate our algorithm for detecting clouds and to classify whether a site is or is not contaminated by ALAN.\\
We analyse the SQM data of the year 2018 at La Silla and at Asiago (Ekar observatory). The measurement with high temporal resolution is carried out every 3 minutes and 5 minutes respectively from 9:00 pm to 5:00 am in local time. We then analyse the data of the GOES satellite\footnote{https://www.class.noaa.gov} as described in Cavazzani et al. (\cite{cava11}) and correlate it with AQUA satellite data using the method described in Cavazzani et al. (\cite{cava15} and \cite{cava17}). 
We correlate each night satellite data with SQM data measured from the ground.\\
At La Silla we have a triple validation of GOES-AQUA-SQM data and a dual validation of AQUA-SQM at Asiago, since this site is outside the GOES field of view.
A first important consideration on which our analysis was based is the detection that in a site not subject to light pollution as La Silla the sky appears darker during the covered nights while it appears brighter in a contaminated site like Asiago. We also note how the trend is made irregular by the presence of clouds in both cases.
For this reason we calculate a standard deviation between three values at La Silla while, in the case of the Ekar observatory, we calculate a maximum half-dispersion between two values due to the lower temporal resolution of the instrument. We have chosen intervals of 9 minutes for La Silla and 10 minutes for Asiago to exclude the gradual variations due to the presence of the Milky Way or the Moon. Our algorithm reproduces the sky brightness and the cloud cover trends during the night. The standard deviation is higher when the night is covered whereas in the case of a clear night it is very low.\\
The conversion of the SQM standard deviation into cloud cover fraction is carried out through the monthly temporal analysis: the time intervals are classified cloudy when the standard deviation is above the threshold function described in Subsection \ref{so} (e.g. We analyze about 240 hours per month, if the standard deviation is above the threshold for 20 of these hours we get $\frac{20}{240}\times100=8.3\%$ of cloud cover fraction).

\begin{figure}
  \centering
  \includegraphics[width=8cm]{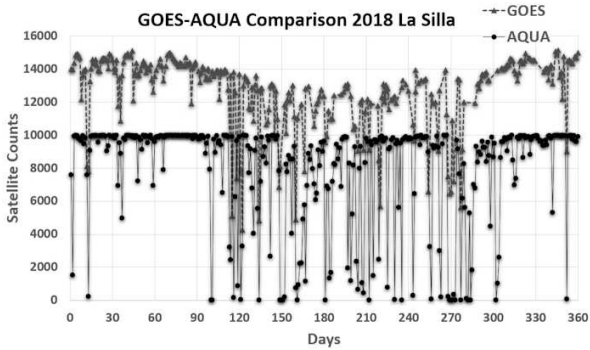}
  \caption{Comparison between GOES and MODIS daily data in 2018 at La Silla. The top trend represents the GOES data and shows the seasonal temperature: the winter months are the coldest and the most covered. The trend of MODIS is instead normalized to the value of $10000$ satellite units.}
             \label{s2018}
\end{figure}

\begin{figure}
  \centering
  \includegraphics[width=8cm]{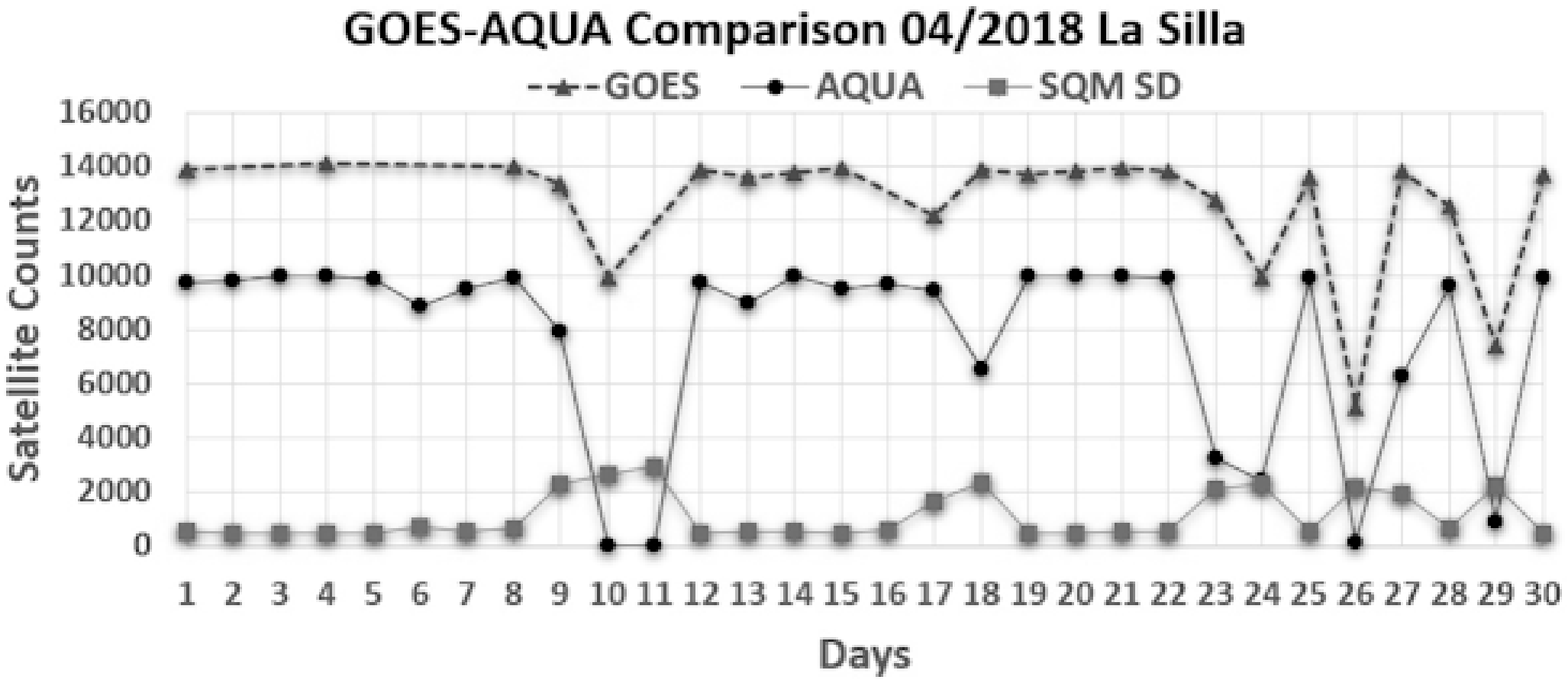}
  \includegraphics[width=4cm]{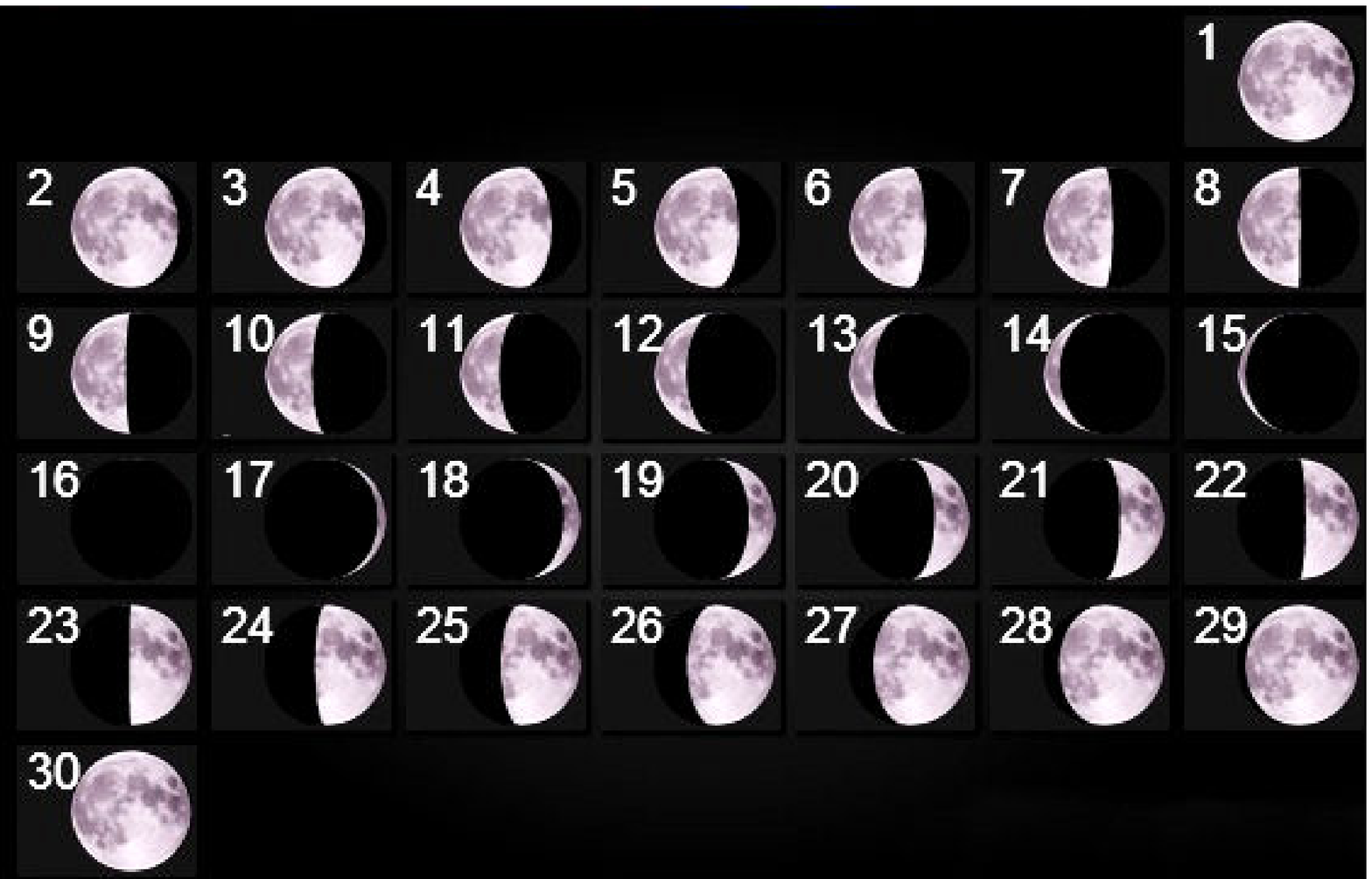}
  \caption{The top panel shows the triple comparison of GOES, MODIS and average standard deviation of SQM data ($SQM_{SD}=\bar{\sigma}\times 10000$). Cloudy nights show a drop in satellite unit count and an increase in average stardard deviation. The bottom panel below shows the month lunar cycle.}
             \label{s04}
\end{figure}

\begin{table}
 \centering
 \begin{minipage}{80mm}
  \caption{Variable thresholds as a function of site magnitude.}
   \label{t}
  \begin{tabular}{@{}lcccccc@{}}
  \hline
Magnitude & $\sigma$ & Magnitude & $\sigma$ \\
\hline

22.0 &	0.050 & 19.5 &	0.164 \\
21.5 &	0.073 & 19.0 &	0.186 \\
21.0 &	0.096 & 18.5 & 	0.209 \\
20.5 &	0.118 & 18.0 & 	0.232 \\
20.0 &	0.141 & 17.5 &  0.255 \\

\hline
\end{tabular}
\end{minipage}
\end{table}

\subsection{Photometric and spectroscopic nights classification using the SQM}
\label{so}

In this Section we describe the empirical-mathematical model used to understand when the value of the standard deviation indicates the presence of clouds.
The choice of the standard deviation threshold is fundamental for the night classification. First of all we have empirically set the threshold as a function of the yearly mean magnitude $M$ (see Equation \ref{mm}) detected by VIIRS (see Figure \ref{map}): 21.9 for La Silla and 20.9 for Ekar observatory.
We assumed a linear relationship between the SQM SD-threshold and the $M$ value detected by satellite, considering that both data families are expressed in magnitude.
This relationship is given by the empirical formula:

\begin{equation}
	\sigma = A\cdot M + B
\label{mm}
\end{equation}

where the values of A and B are empirically obtained through the correlation with the GOES and AQUA satellite data. We assumed a minimum threshold value of 0.050 for a site with a magnitude of 21.9 ($0.050 = A\cdot 21.9 + B$) and a value of 0.100 for a site with magnitude 20.9 ($0.100 = A\cdot 20.9 + B$). We calculate the values in Table \ref{t} by solving a linear system and obtaining A=-0.04545 and B=1.05000.\\
Note that the algorithm uses a lower threshold for sites less contaminated by ALAN, in agreement with Figure 8 published in Puschnig et al. (\cite{pus}).\\
The second factor that influences the threshold is the presence of the Moon, as shown in Figure \ref{ath}. 
We argue that, in a site that is not contaminated by ALAN, and close to the new Moon, the clouds block the natural contribution of the sky, magnified the fluctuations detected by the SQM. Therefore our algorithm uses a higher threshold during these days. 
During the nights near the full Moon, the clouds are illuminated only from above, and therefore they dampen the fluctuations encountered by the SQM, making the sky more homogeneous.
The opposite occurs in a site contaminated by ALAN. The clouds are strongly illuminated from below and this in days without the presence of the Moon eliminates all the fluctuations due to the natural light of the sky lowering the SQM standard deviation. 
During the nights near the full Moon, the clouds are instead illuminated from above and from below, increasing the SQM fluctuations. Therefore, we suggest to approximate the threshold by the following function for a site not contaminated by ALAN:

\begin{equation}
	\sigma (x)= \frac{\Lambda}{2}\cdot cos\left[\frac{2\pi}{T}(x)\right]+ \sigma
\end{equation}

where $\Lambda$ is minimum threshold value, $x$ are the days starting from the new Moon, $\sigma$ is provided by the Table \ref{t} and T is the lunar synodic period chosen for the classification.
The function for a contaminated site becomes:

	\[ \sigma (x)= -\frac{\Lambda}{2}\cdot cos\left[\frac{2\pi}{T}(x)\right]+ \sigma
\]

\begin{figure}
  \centering
  \includegraphics[width=8cm]{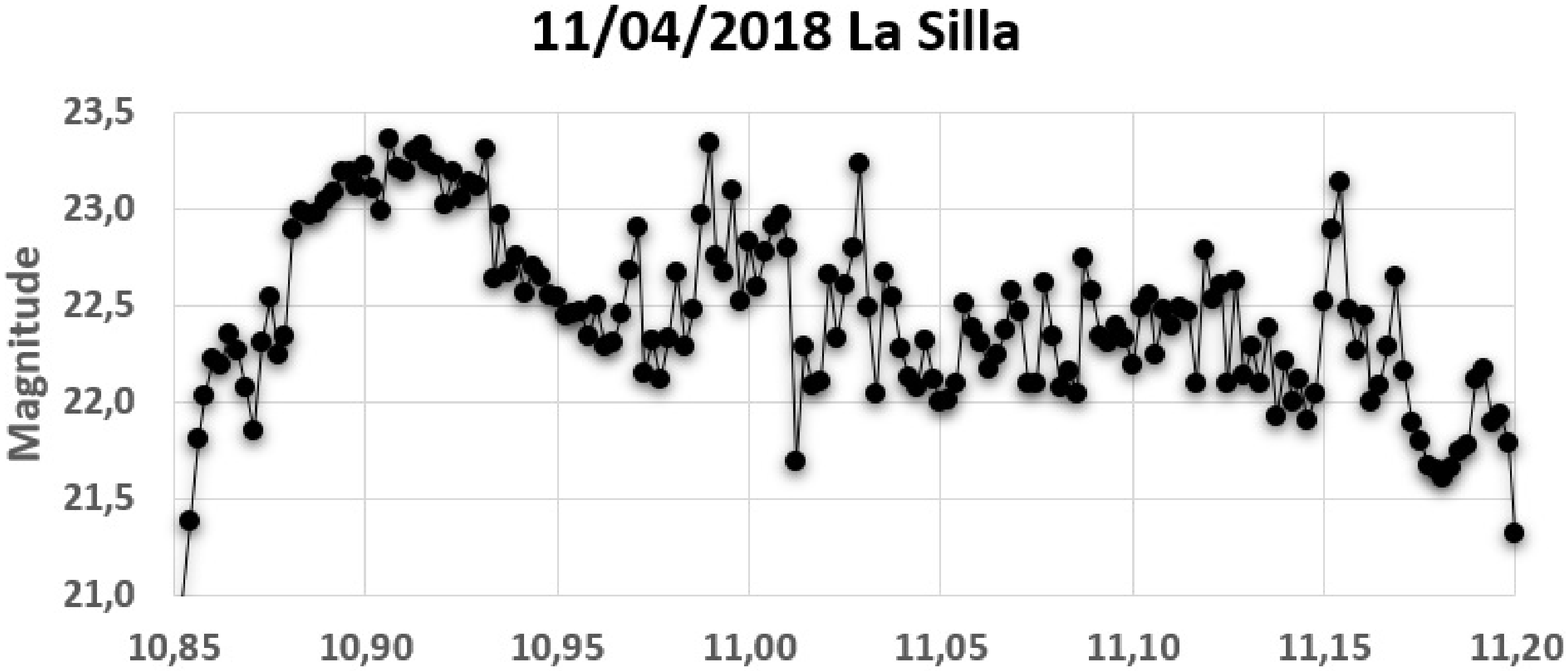}
  \includegraphics[width=8cm]{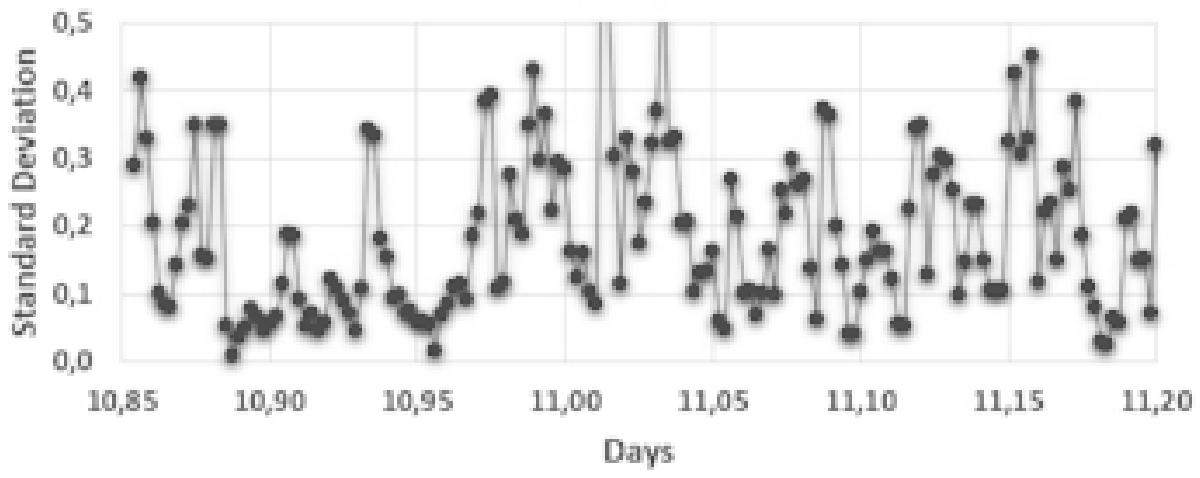}
  \caption{Trend of SQM values in a night with cloudy sky in last quarter of Moon at La Silla, 10 April 2018 (top panel) and the respective standard deviation trend every 9 minutes (bottom panel). The SQM values and SD are expressed in $mpsas=\frac{mag}{arcsec^{2}}$ (y-axis).}
             \label{s10}
\end{figure}

\begin{figure}
  \centering
  \includegraphics[width=8cm]{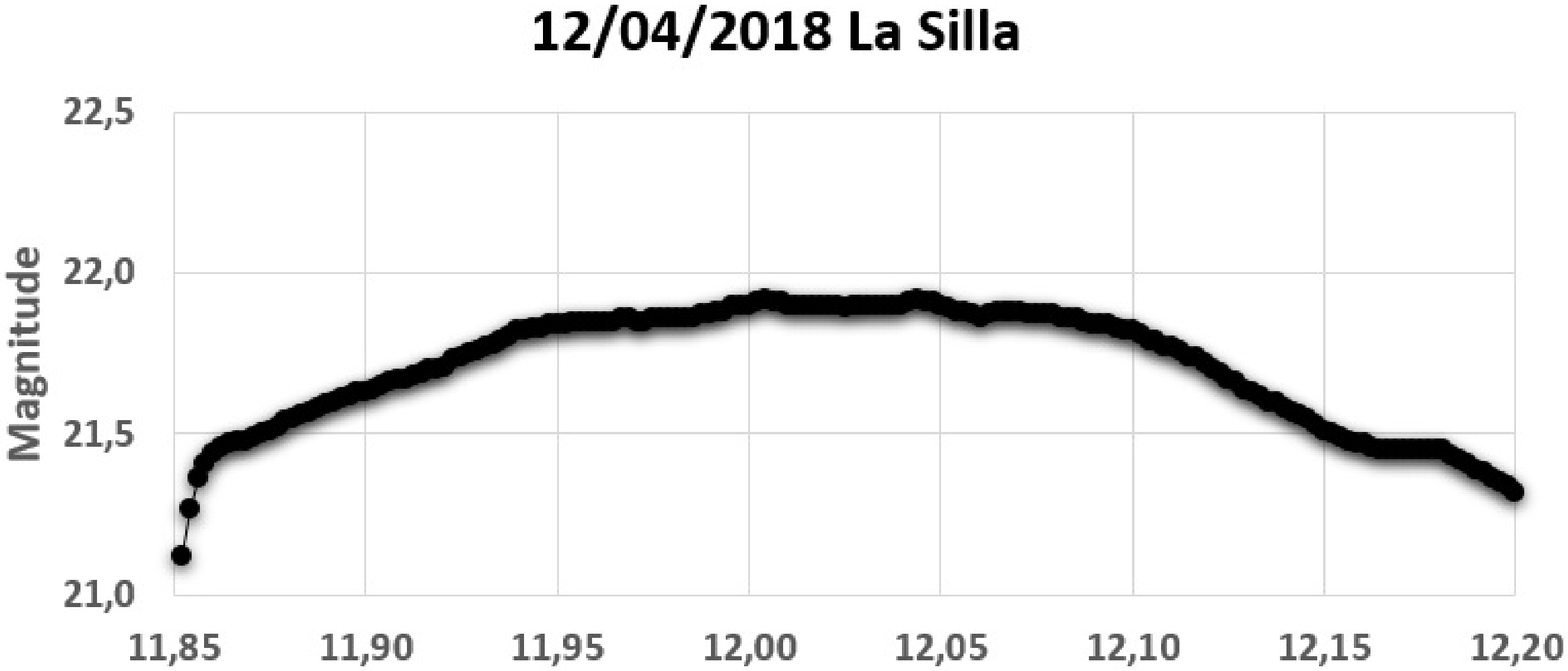}
  \includegraphics[width=8cm]{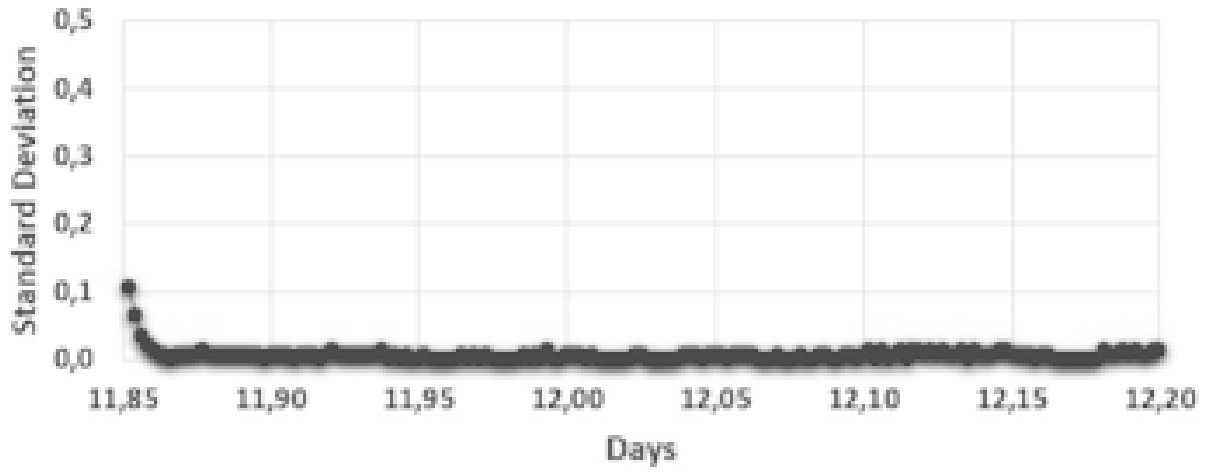}
  \caption{Trend of SQM values in a night with clear sky in last quarter of Moon at La Silla, 12 April 2018 (top panel) and the respective standard deviation trend every 9 minutes (bottom panel). The SQM values and SD are expressed in $mpsas=\frac{mag}{arcsec^{2}}$ (y-axis).}
             \label{s12}
\end{figure}

\begin{figure}
  \centering
  \includegraphics[width=8cm]{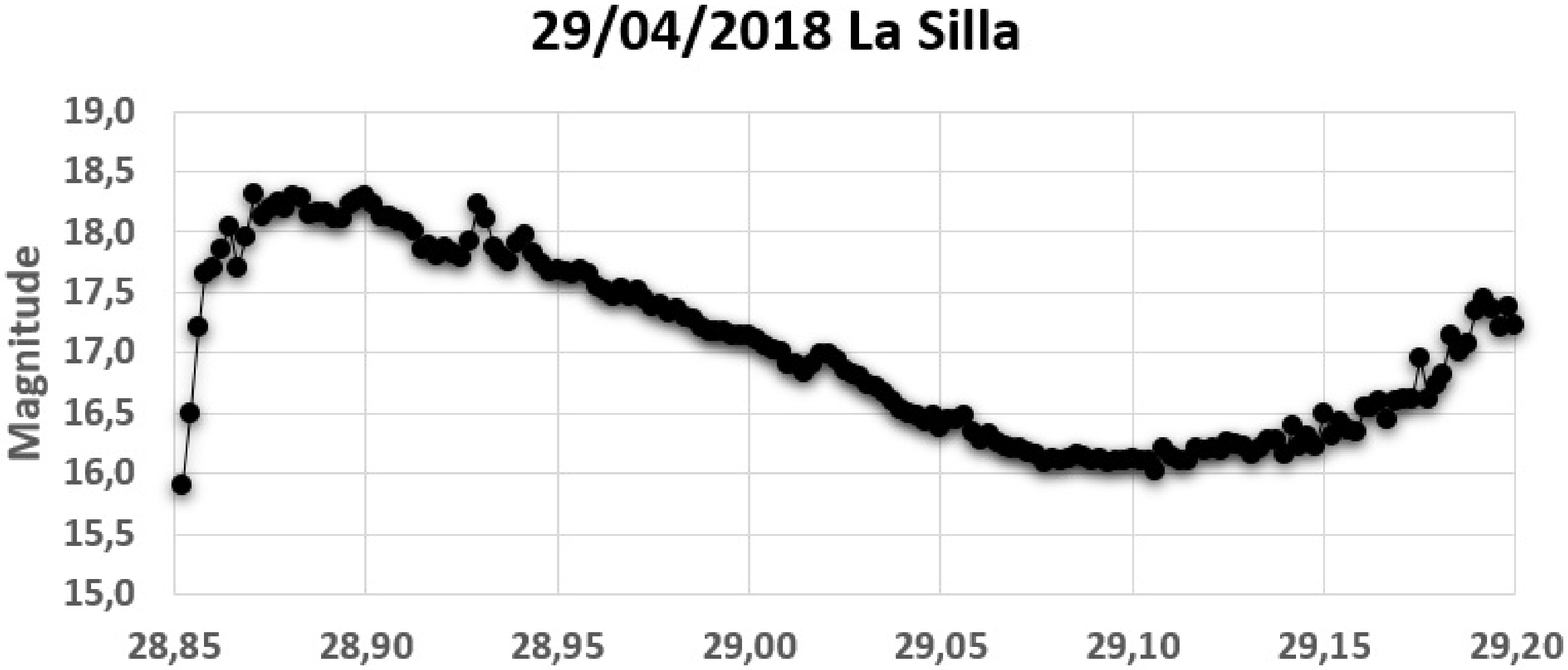}
  \includegraphics[width=8cm]{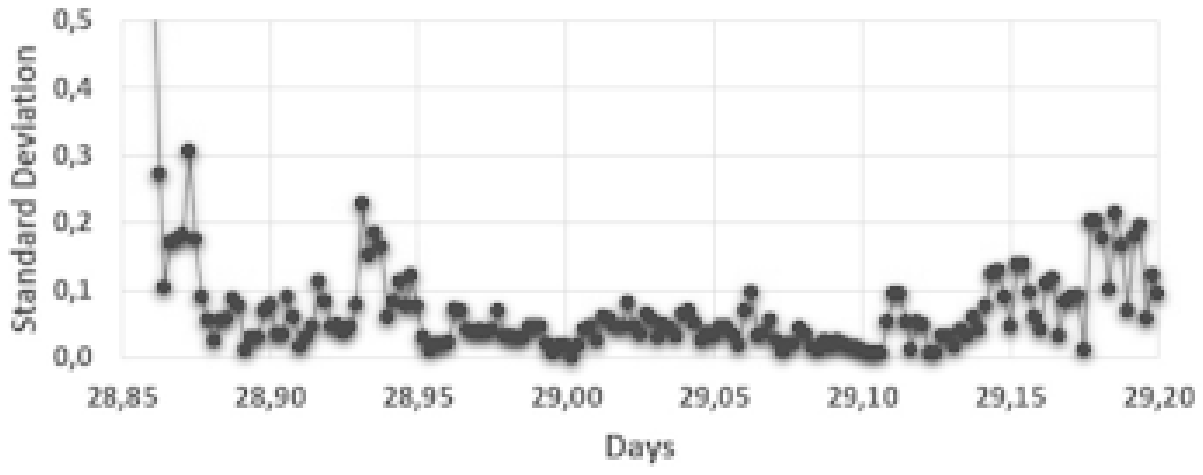}
  \caption{Trend of SQM values in a night with cloudy sky near the full Moon at La Silla, 28 April 2018 (top panel) and the respective standard deviation trend every 9 minutes (bottom panel). The SQM values and SD are expressed in $mpsas=\frac{mag}{arcsec^{2}}$ (y-axis).}
             \label{s28}
\end{figure}

\begin{figure}
  \centering
  \includegraphics[width=8cm]{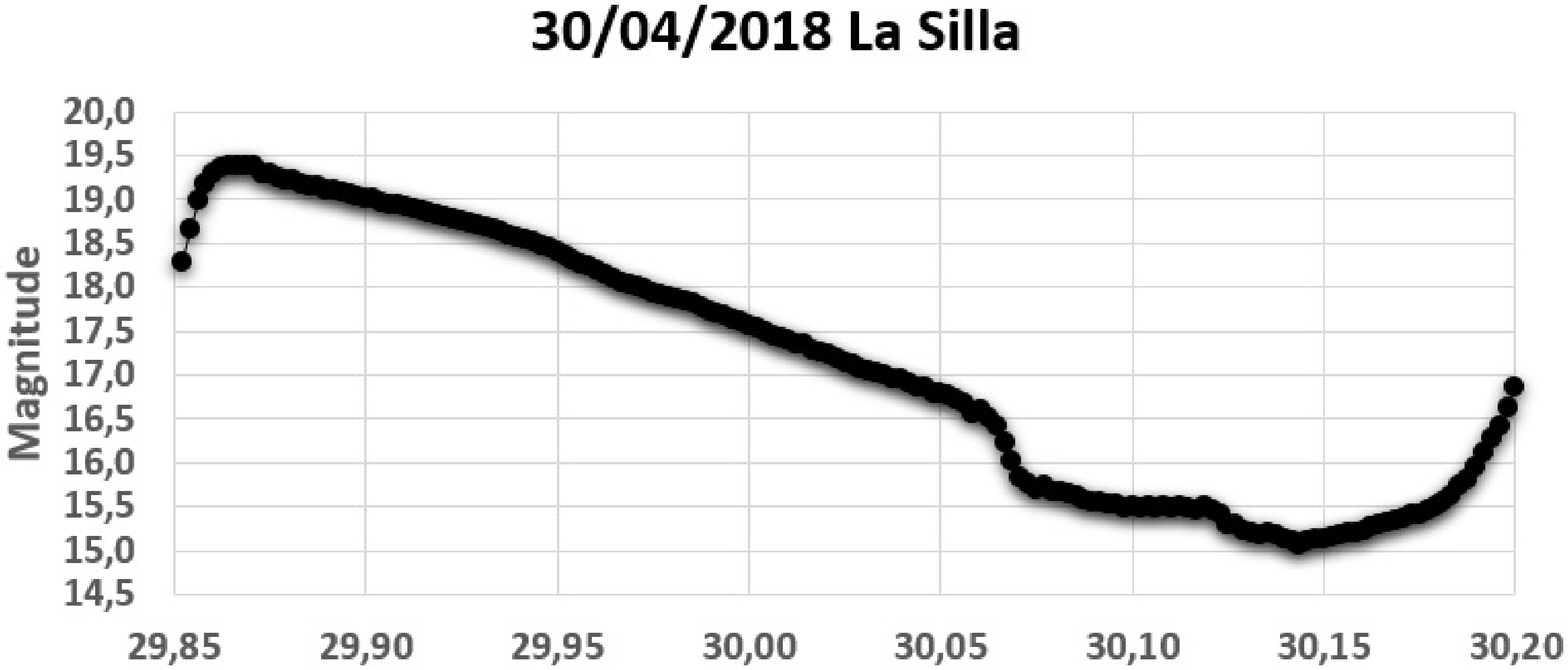}
  \includegraphics[width=8cm]{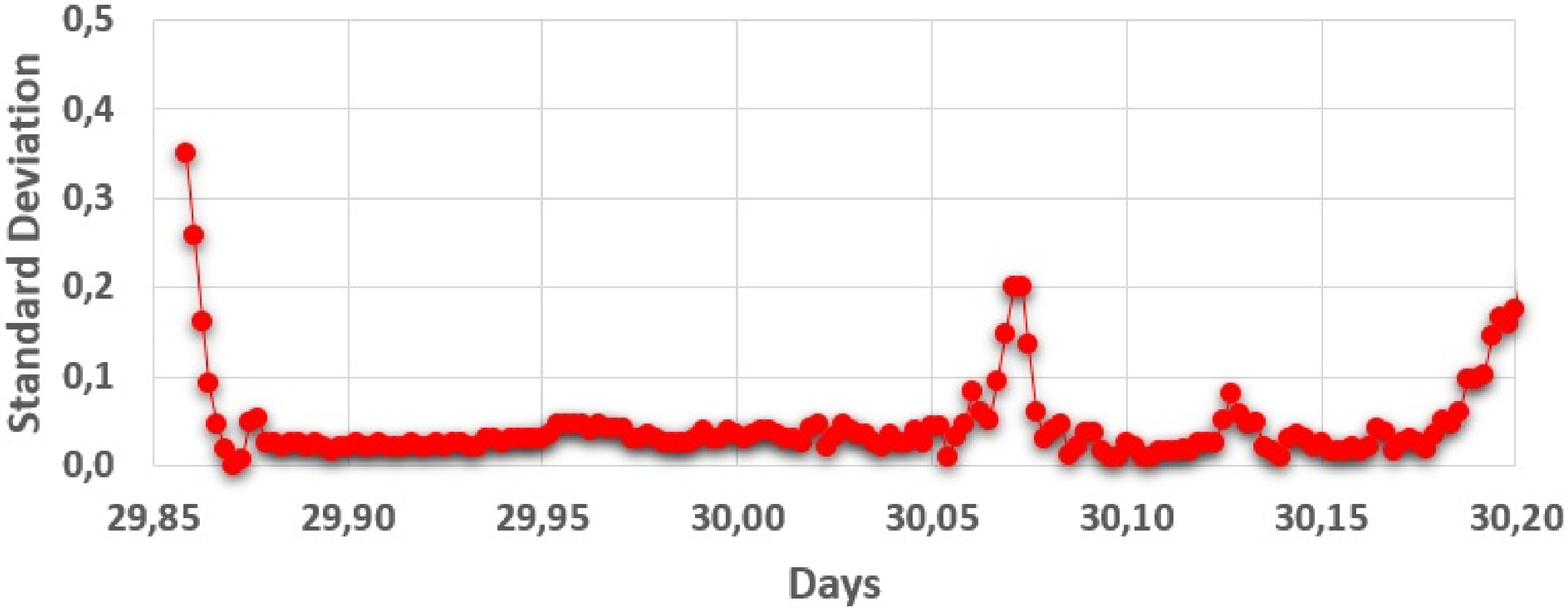}
  \caption{Trend of SQM values in a night with clear sky near the full Moon at La Silla, 29 April 2018 (top panel) and the respective standard deviation trend every 9 minutes (bottom panel). The SQM values and SD are expressed in $mpsas=\frac{mag}{arcsec^{2}}$ (y-axis).}
             \label{s29}
\end{figure}

We classify the nights through this function following the definition of photometric night (PN) and spectroscopic night (SN). We consider PN the nights with an interval greater than 6 hours under the threshold function, while SP with a interval greater of 2 hours. We analyse local time from 9:00 pm to 5:00 am for a total of 4800 monthly data (N=160 per night) at La Silla and 2280 at Asiago (N=96 per night).

\section{La Silla observatory}
\label{c}

Figure \ref{s2018} shows the comparison between the GOES and the MODIS data in 2018 at La Silla. The cloud cover detected is 81.3\% and 80.1\%, respectively. The punctual correlation between the two satellite data sets, calculated with the Pearson correlation index, for this site is 96.6\%.\\
Figure \ref{s04} shows in detail the month of April 2018 (top panel) with the respective lunar cycle (bottom panel). Choosing this month allowed us to analyse four basic cases: clear or cloudy sky in nights near the new Moon and clear or cloudy sky near the full Moon. The top trend represents the GOES data, the central trend represents the MODIS data and the bottom trend at is the average standard deviation of the SQM data every 9 minutes. The MODIS data are analysed as in Cavazzani et al. (\cite{cava15}) for comparison with GOES data, while we used the following conversion to compare the mean standard deviation with the satellite data:

	\[SQM_{SD}=\bar{\sigma}\times 10000
\]

\begin{figure}
  \centering
  \includegraphics[width=8cm]{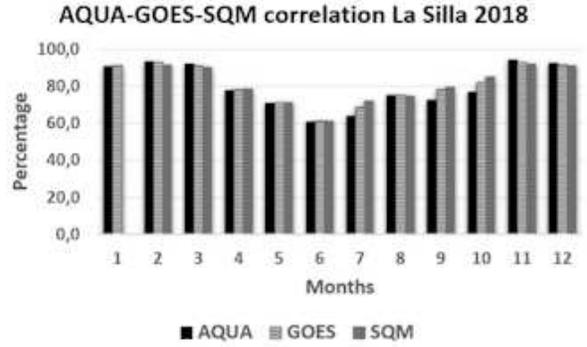}
  \caption{Comparison of the clear sky percentages (y-axis) measured with AQUA, GOES and the SQM standard deviation in 2018 at La Silla.}
             \label{lscor}
\end{figure}

\begin{figure}
  \centering
  \includegraphics[width=8cm]{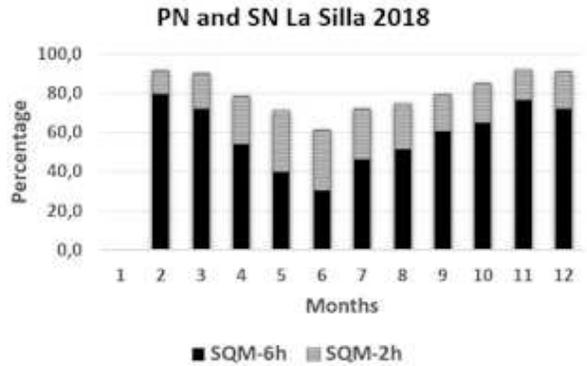}
  \caption{Photometric and spectroscopic nights calculated by the SQM algorithm for cloud detection in 2018 at La Silla.}
             \label{lspn}
\end{figure}

We see a low standard deviation in correspondence with clear sky conditions and vice versa. We analyse in detail the nights with the four conditions described above: Figure \ref{s10} shows the trend of a covered night in a day of last Moon quarter (top panel). The presence of clouds makes the sky darker in a site not contaminated by ALAN. The magnitude value is greater than 23, and remains about 22 for the whole night, with large fluctuations. The bottom panel shows the respective standard deviation calculated every 9 minutes, the average value is 0.33.
Figure \ref{s12} shows the SQM data of a last  quarter clear night. During this night the maximum value is 21.8 while the average standard deviation is 0.03. Figure \ref{s28} shows the status of a covered night near full Moon. The maximum value is 18.2 while the minimum is 16.0, the average standard deviation is 0.18. Finally Figure \ref{s29} shows the trend of a full Moon clear night, the maximum value is 19.4 while the minimum is 15.0. The average standard deviation is 0.05. By comparison of the Figures \ref{s28} and \ref{s29} we see that in a site without ALAN with the presence of the Moon, the clouds make the sky brighter in the first part of the night and darker in the central part.\\
Table \ref{t1} shows the results obtained in terms of clear sky time. This type of analysis provides the sum of photometric and spectroscopic observation time in a site with very stable night time conditions: in particular, a night that begins with a clear sky, remains good (Cavazzani et al., \cite{cava12}). Column 2 shows the clear sky percentage detected by AQUA, column 3 by GOES (see Section \ref{aq}) and column 4 by the standard deviation of SQM data. Column 5 shows the daily correlation between the two satellites (A-G) while column 6 between GOES and the SQM standard deviation (G-SQM). The last line gives the annual average values.
The comparison between the results provided by GOES, AQUA and our algorithm are shown in Figure \ref{lscor}. Figure \ref{lspn} and Table \ref{t2} show the percentages of PN and SP at La Silla in 2018 calculated with the SQM standard deviation.
We carried out a further check using a sample, one month of ground data\footnote{http://archive.eso.org/cms/eso-data/ambient-conditions.html}. We have chosen the month of May for its climatic complexity in order to verify the algorithm in various climatic conditions. We measure a PN percentage of 38.7\% and a SN of 29.0\% from the ground data with a SQM punctual correlation on the single night of 96.3\%.

\begin{figure}
  \centering
  \includegraphics[width=8cm]{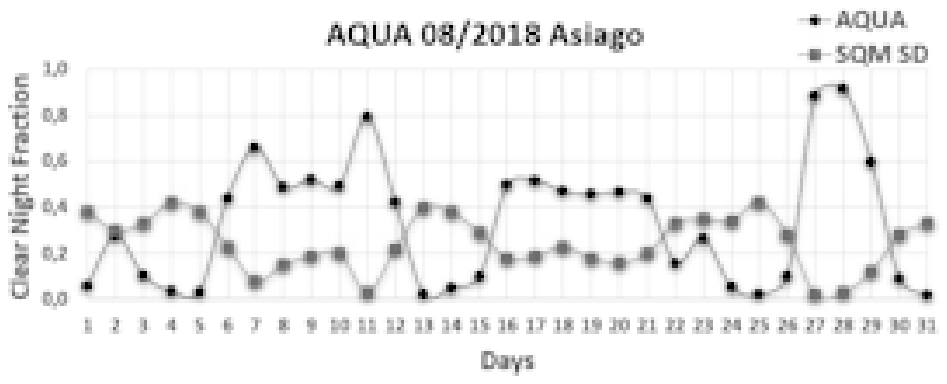}
  \includegraphics[width=4cm]{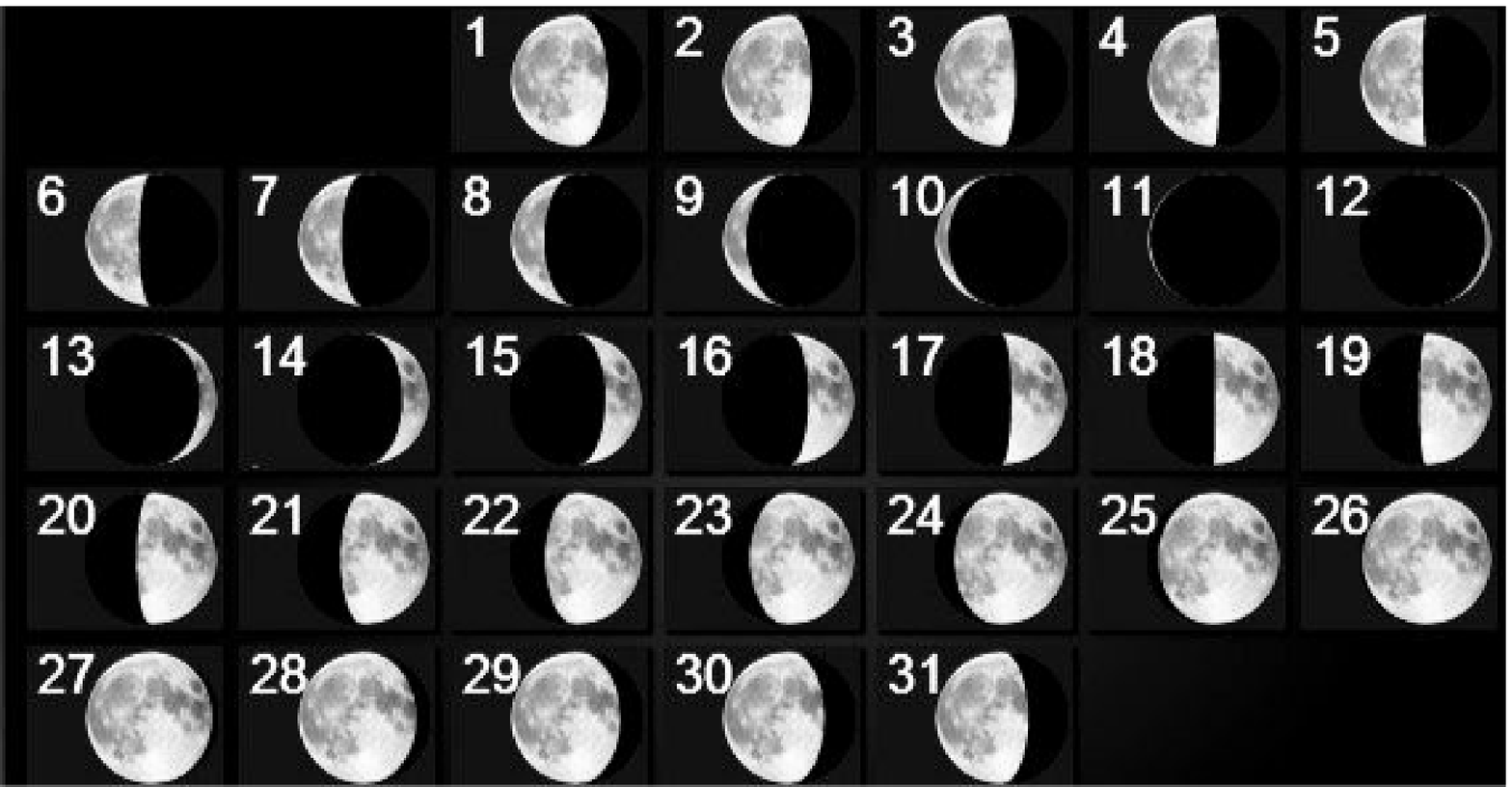}
  \caption{The top panel shows the comparison between the MODIS and the average maximum half-dispersion of SQM data ($SQM_{SD}=\bar{\sigma}\times 10000$). Cloudy nights show a drop in satellite unit count and an increase in average maximum half-dispersion. The bottom panel below shows the month lunar cycle.}
             \label{a08}
\end{figure}

\begin{figure}
  \centering
  \includegraphics[width=8cm]{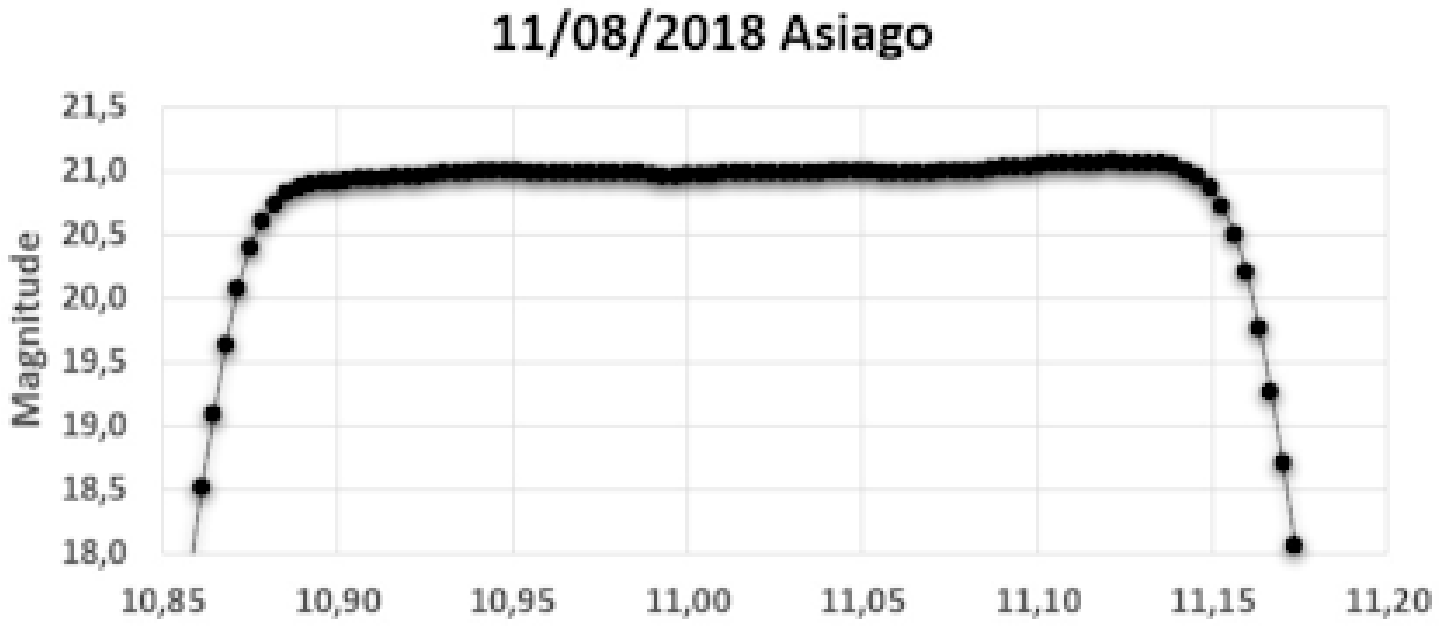}
  \includegraphics[width=8cm]{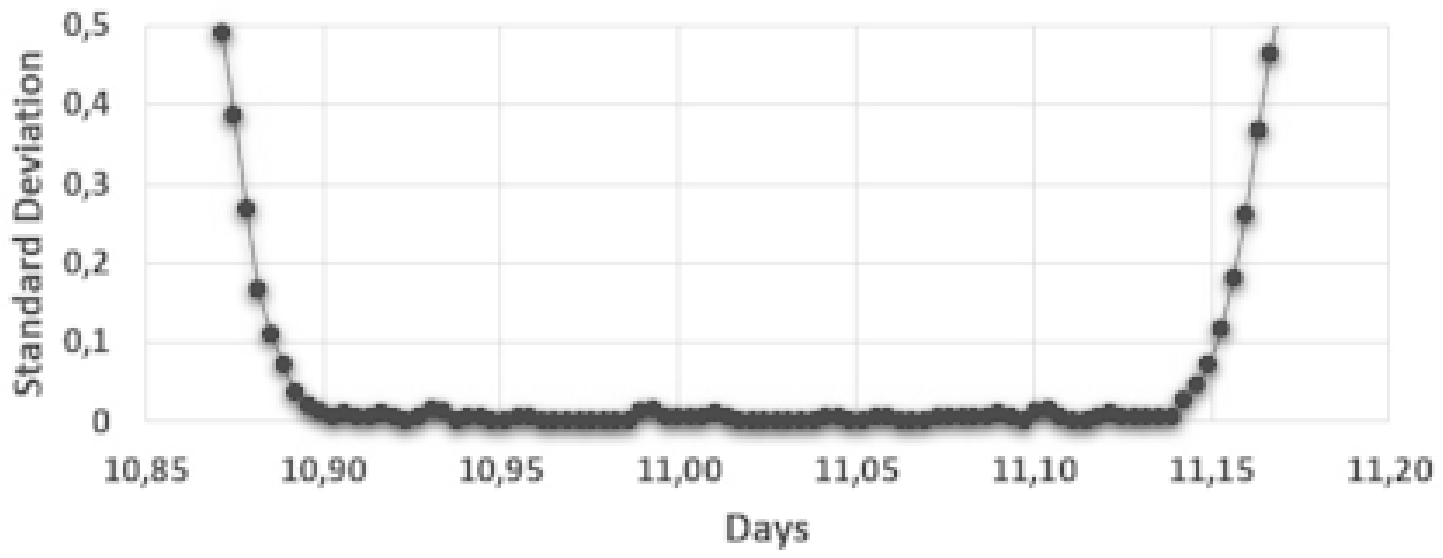}
  \caption{Trend of SQM values in a night with new Moon clear sky at Ekar Observatory, 10 August 2018 (top panel) and the respective maximum half-dispersion trend every 10 minutes (bottom panel). The SQM values and maximum half-dispersion are expressed in $mpsas=\frac{mag}{arcsec^{2}}$ (y-axis).}
             \label{a11}
\end{figure}

\begin{figure}
  \centering
  \includegraphics[width=8cm]{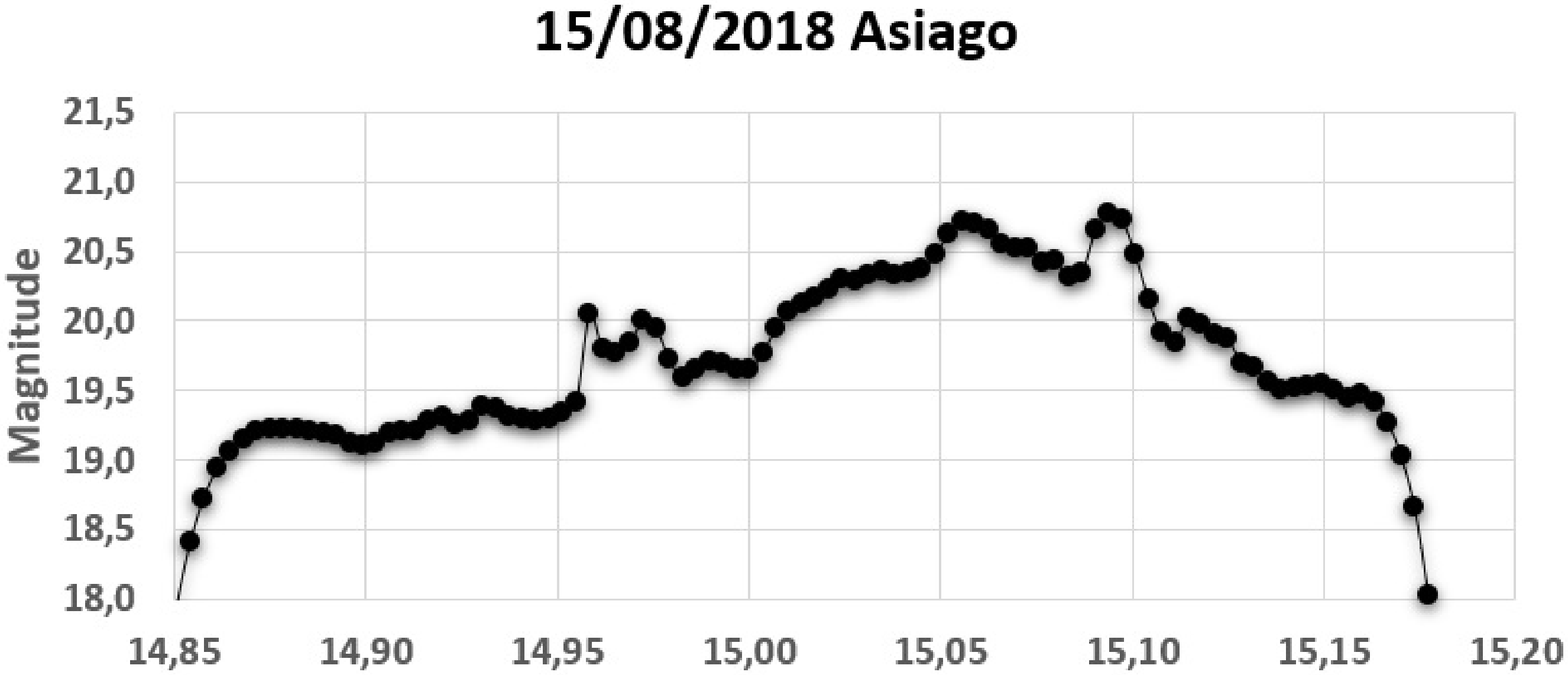}
  \includegraphics[width=8cm]{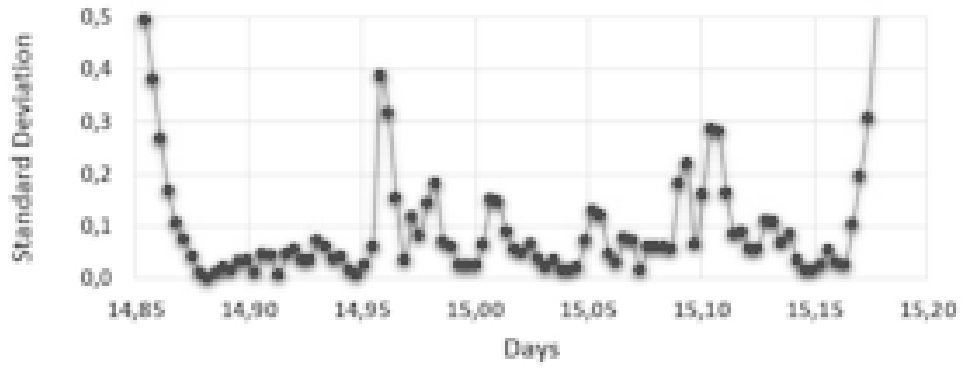}
  \caption{Trend of SQM values in a night with first Moon quarter cloudy sky at Ekar Observatory, 15 August 2018 (top panel) and the respective maximum half-dispersion trend every 10 minutes (bottom panel). The SQM values and maximum half-dispersion are expressed in $mpsas=\frac{mag}{arcsec^{2}}$ (y-axis).}
             \label{a15}
\end{figure}

\begin{figure}
  \centering
  \includegraphics[width=8cm]{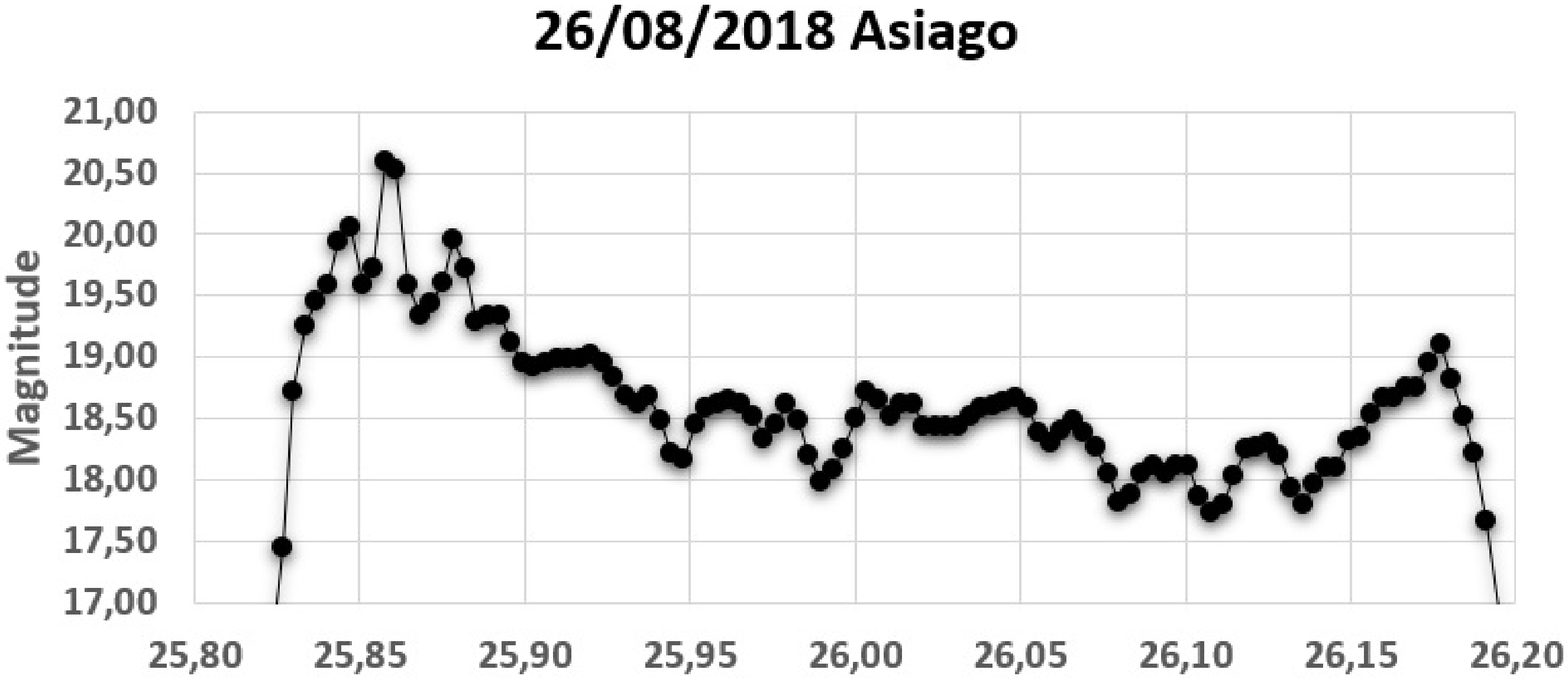}
  \includegraphics[width=8cm]{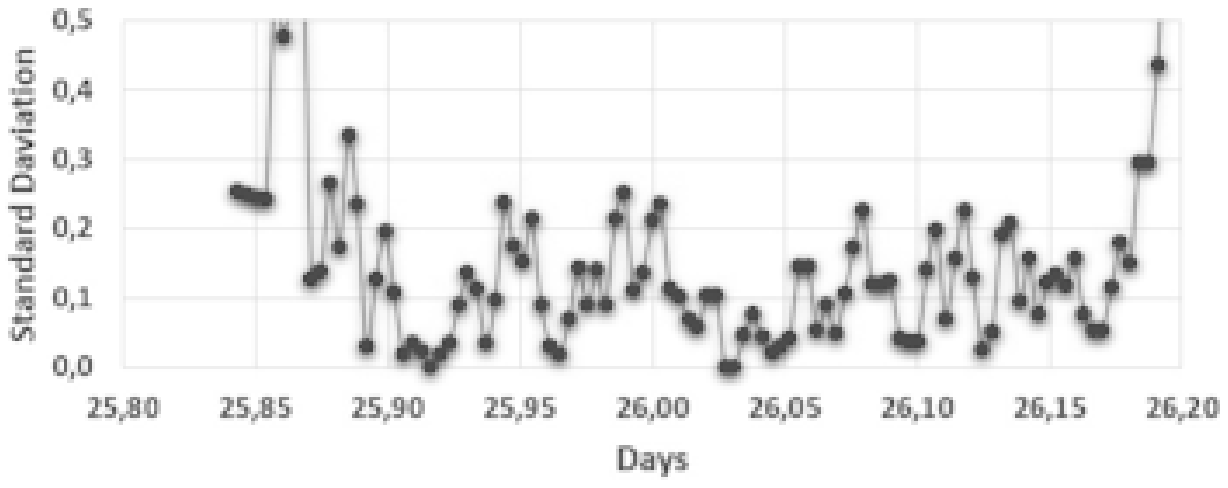}
  \caption{Trend of SQM values in a night with full Moon cloudy sky at Ekar Observatory, 26 August 2018 (top panel) and the respective maximum half-dispersion trend every 10 minutes (bottom panel). The SQM values and maximum half-dispersion are expressed in $mpsas=\frac{mag}{arcsec^{2}}$ (y-axis).}
             \label{a26}
\end{figure}

\begin{figure}
  \centering
  \includegraphics[width=8cm]{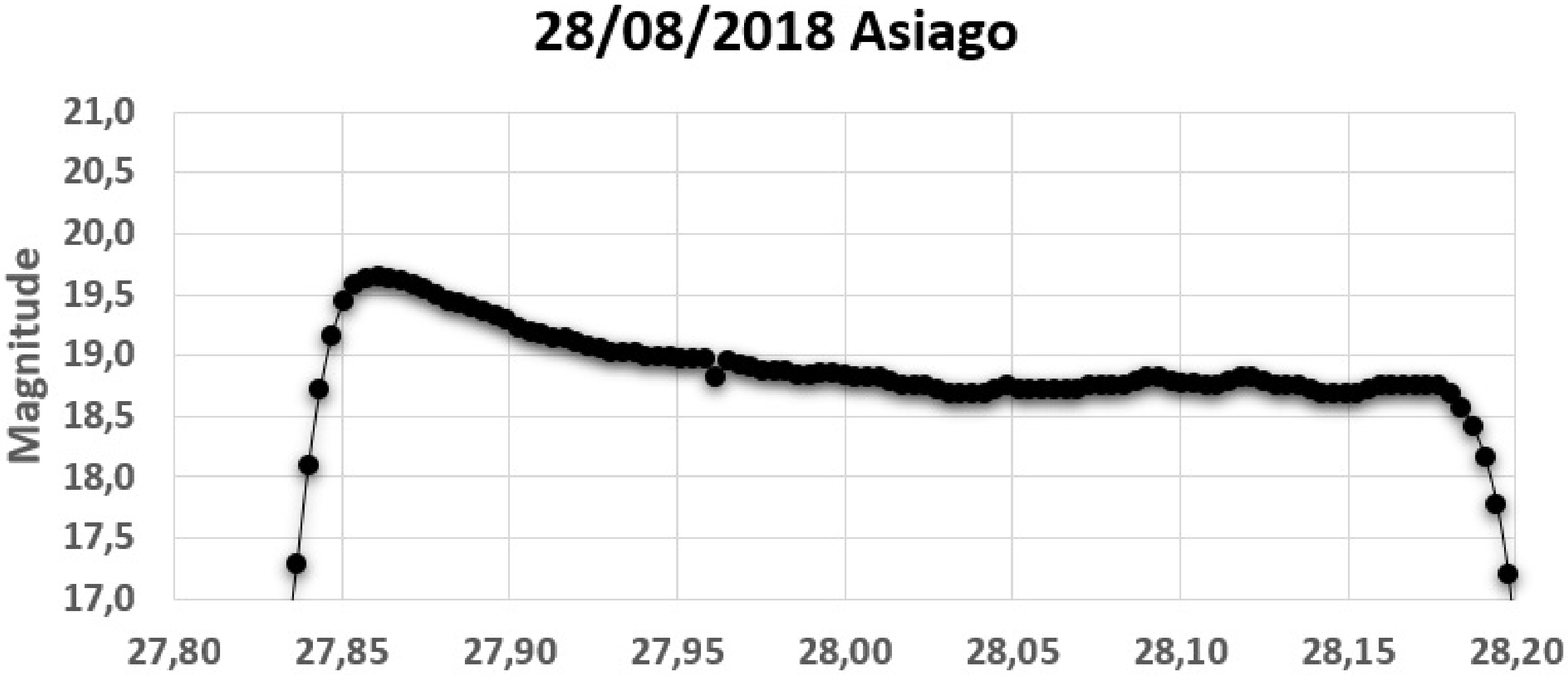}
  \includegraphics[width=8cm]{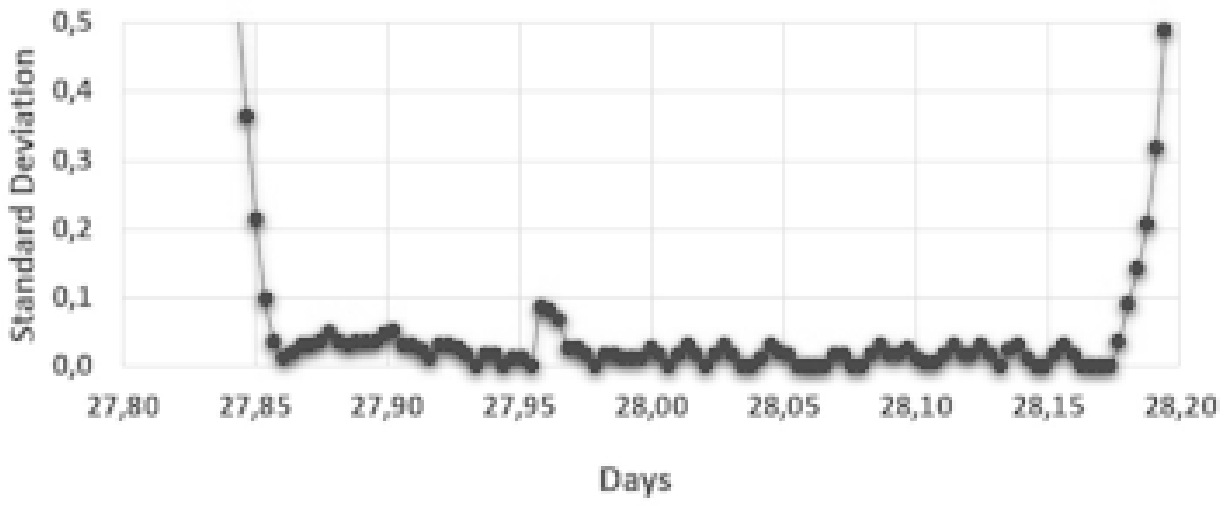}
  \caption{Trend of SQM values in a night with almost full Moon clear sky at Ekar Observatory, 28 August 2018 (top panel) and the respective maximum half-dispersion trend every 10 minutes (bottom panel). The SQM values and maximum half-dispersion are expressed in $mpsas=\frac{mag}{arcsec^{2}}$ (y-axis).}
             \label{a28}
\end{figure}

\begin{table}
 \centering
 \begin{minipage}{80mm}
  \caption{Night clear sky percentage at la Silla in 2018. Triple daily correlation between GOES and AQUA (A-G) satellite and the SQM algorithm for the night clouds detection (G-SQM). Column 2 shows the clear sky percentage detected by AQUA, column 3 by GOES and column 4 by the standard deviation of SQM data. Column 5 shows the daily correlation between the two satellites (A-G) while column 6 between GOES and the SQM standard deviation (G-SQM). The last line gives the annual average values.}
   \label{t1}
  \begin{tabular}{@{}lccccccccccc@{}}
  \hline

La Silla  &\multicolumn{4}{c}{GOES-ACQUA-SQM correlations}		\\ 
  \hline
2018 &	AQUA &	GOES &	SQM & 	A-G &	G-SQM \\
\hline
1	& 90.9 &	91.3 &	- &	    99.2 &	- \\   
2	& 93.3 &	92.8	& 91.8 & 	99.0 &	97.5 \\
3 &	92.2 &	91.3 &	90.2 &	98.2& 	97.3 \\
4	& 77.9 &	78.4 &	78.6 &	99.0 &	99.5 \\
5	& 70.9 &	71.3 &	71.4 &	99.2 &	99.8 \\
6 &	60.9 &	61.3 &	61.5 &	99.2 &	99.5\\
7	& 64.1 &	68.6 &	72.4 &	91.0 &	90.5 \\
8	& 75.2 & 75.3	& 74.8 & 	99.8 &  	98.8 \\
9	& 72.7 &	78.3 &	79.4 &	88.8 &	97.3 \\
10 & 76.8	& 82.1 & 	85.0 &	89.4 &	92.8 \\
11 &94.2 &	92.9 &	91.9 &	97.4 &	97.5 \\
12 &92.6 &	91.8 &	91.4 &	98.4 &	99.0 \\
\hline
Mean& 80.1 &	81.3 &	80.8 &	96.6 &	97.2 \\
\hline
\end{tabular}
\end{minipage}
\end{table}

\begin{table}
 \centering
 \begin{minipage}{80mm}
  \caption{Night clear sky percentage at la Silla in 2018. Percentage of photomentric (SQM-6h) and spectroscopic (SQM-2h) nights.}
   \label{t2}
  \begin{tabular}{@{}lccccccccccc@{}}
  \hline

2018 &	SQM-6h &	SQM-2h	& SQM \\
\hline
1 &	-	& - & - \\
2	& 79.4 &	12.4 &	91.8 \\
3	 & 72.1	 & 18.1 & 	90.2 \\
4 &	54.3 &	24.3 &	78.6 \\
5 &	40.1 &	31.3 &	71.4 \\
6	& 30.7	& 30.8 &	61.5 \\
7 &	46.6	& 25.8 &	72.4 \\
8 &	51.4 &	23.4 &	74.8 \\
9	& 61.1 &	18.3 &	79.4 \\
10 &	65.2	& 19.8 &	85.0 \\
11	& 76.6 &	15.3	& 91.9 \\
12	& 72.1 &	19.3 &	91.4 \\
\hline
Mean	& 59.1 &	21.7	& 80.8 \\
\hline
\end{tabular}
\end{minipage}
\end{table}

\section{Ekar Observatory in Asiago}
\label{d}

\begin{figure}
  \centering
  \includegraphics[width=8cm]{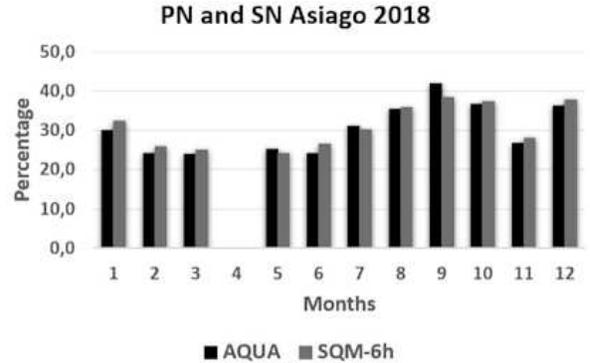}
  \caption{{Comparison of the clear sky percentages (y-axis) measured with} AQUA and the SQM maximum half-dispersion in 2018 at Ekar observatory.}
             \label{acor}
\end{figure}

\begin{figure}
  \centering
  \includegraphics[width=8cm]{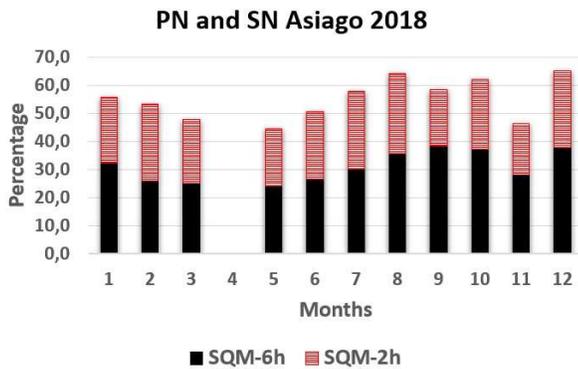}
  \caption{Photometric and spectroscopic nights calculated by the SQM algorithm for cloud detection in 2018 at Ekar observatory.}
             \label{apn}
\end{figure}

We perfomed the same analysis for the Ekar observatory in Asiago. Figure \ref{a08} shows the correlation between the mean maximum half-dispersion and the AQUA data at Asiago in August 2018. We analyse in detail the four main conditions that can be found on a site. The top panel of Figure \ref{a11} shows a clear night of new Moon with its relative maximum half-dispersion (bottom panel). The night reaches magnitude values greater than 21. Top panel of Figure \ref{a15} shows a a first quarter Moon night with clouds. This shows that in a contaminated site the clouds decrease the magnitude of the sky. The bottom panel shows the SQM maximum half-dispersion. We can see an example of a spectroscopic night in its first part. 
Top panel of Figure \ref{a26} shows a full Moon covered night. The lower panel shows how the presence of clouds illuminated from above and below increases the SQM maximum half-dispersion. This explains the use of the variable lunar threshold shown in Figure \ref{ath}.
Finally, Figure \ref{a28} shows a clear night next to the full Moon with its respective maximum half-dispersion.
Asiago is a site highly contaminated by ALAN (see Figure \ref{map}) and also has high cloud cover conditions, around 40\% per year\footnote{http://www.oapd.inaf.it}. Finally it does not have stable night time conditions, therefore the low temporal and spatial resolution of the AQUA satellite does not provide results in terms of observation time but in terms of photometric night. The algorithm classifies intervals longer than 6 hours as photometric nights and intervals longer than 2 hours as spectroscopic nights.\\
Table \ref{t3} shows the obtained results: column 2 gives the AQUA satellite values and column 3 the values of the photometric nights detected by the SQM. Column 4 shows the monthly point correlation between the two groups of data (S-SQM-6h). Column 5 shows the added percentage by spectroscopic nights. The sum of the columns 4 and 5 gives the annual percentage of site use. The last line gives the annual average values. Figure \ref{acor} shows the comparison between the photometric nights detected by satellite and those calculated by our algorithm. Figure \ref{apn} shows the sum of the monthly averages of the photometric and spectroscopic nights.
We carried out the same further check with a sample month of ground data\footnote{http://www.oapd.inaf.it} made at La Silla. We have chosen the month of September for its climatic complexity in order to verify the algorithm in various climatic conditions. We measure a PN percentage 36.7\% and a SN 20.0\% from the ground data with a SQM punctual correlation on the single night of 94.1\%. In some cases we found a discrepancy between the ground used time and the SQM data due to high humidity with clear sky conditions.
We made a further analysis at Asiago due to the instability of the site. We calculated the usable time in the first part (20:00-01:00) and in the second part (01:00-06:00) of the night. 
This can be useful in sites with a high percentage of cloud cover to understand the statistically better part of the night for the observation.\\  
Table \ref{t4} shows the results of this analysis: the second part of the night is statistically better at Asiago with the highest discrepancy during the springs in conjunction with the melting of the snow.

\begin{table}
 \centering
 \begin{minipage}{80mm}
  \caption{Night clear sky at Asiago in 2018. Daily correlation between AQUA satellite and the SQM algorithm for the night clouds detection (Column 4, S-SQM-6h). Table shows the clear sky percentage: columns 2 and 3 show the percentages of photometric nights, while column 5 gives the percentage of spectroscopic observation time at least 2-hours intervals.}
   \label{t3}
  \begin{tabular}{@{}lccccccccccc@{}}
  \hline

Asiago  &\multicolumn{4}{c}{ACQUA-SQM correlations}		\\ 
  \hline
2018	& AQUA &	SQM-6h &	S-SQM-6h &	SQM-2h \\
\hline
1 &	30.1	& 32.5 &	91.6  &	23.2  \\
2	 & 24.2	 & 25.9  &	94.1 &	27.3  \\
3  &	23.9	& 25.1	 & 95.8	 & 22.6  \\
4		&   - & - & - & - \\		
5	 & 25.2 &	24.1 &	96.2 &	20.5 \\
6	 & 24.3	 & 26.5	 & 92.3 & 	24.1 \\
7 &	31.2 &	30.2 &	96.5 &	27.6 \\
8	 & 35.4	& 35.8 &	98.6 &	28.5 \\
9	 & 42.0 & 	38.5 &	87.8 &	19.8 \\
10	& 36.7 & 	37.3 &	97.9 & 	24.9 \\
11 &	26.7 &	28.1 &	95.1 &	18.2 \\
12 &	36.4 &	37.9 &	94.8 &	27.3 \\
\hline
Mean &	30.6 &	31.1 &	94.6 &	24.0\\
\hline
\end{tabular}
\end{minipage}
\end{table}

\begin{table}
 \centering
 \begin{minipage}{80mm}
  \caption{Night clear sky at Asiago in 2018 obtained through the SQM maximum half-dispersion. Column 2 gives the percentage of spectroscopic observation time at least 2-hours intervals in the first part of the night (20:00-01:00), while column 3 in the second part of the night (01:00-06:00).}
   \label{t4}
  \begin{tabular}{@{}lcccccc@{}}
  \hline

Month	  &	SQM-2h  & SQM-2h  \\
&    (20:00-01:00) &  (01:00-06:00)  \\
\hline
January	  & 52  & 55  \\
February 	& 53 & 55	\\
March    	& 45 & 53	\\
April	    & 46 & 56 \\
May       & 47 & 50 \\
June      & 57 & 60 \\
July    	& 65  & 69 \\
August    & 63  & 68 \\
September	&  58 & 60 \\
October	  &  53 & 58 \\
November  &  47 & 52 \\
December  &	53  & 57 \\
\hline
Mean      & 53 & 58  \\
\hline
\end{tabular}
\end{minipage}
\end{table}

\section{Correlation uncertainty between the SQM and satellite data}
\label{error}

We associated two types of uncertainties with our correlation measurement: a monthly statistical uncertainty and a punctual nightly uncertainty. 
Table \ref{t1} shows the monthly night clear sky percentages obtained by satellite and the SQM readings at La Silla. 
We consider the monthly statistical uncertainty between the two satellites given by the absolute value of the difference between column 2 and column 3 while between GOES and SQM the absolute value of the difference between column 3 and column 4. The annual averages are $\epsilon_{A-G}=0.2\%$ and $\epsilon_{G-SQM}=0.5\%$ respectively at La Silla in 2018. 
The punctual uncertainty, as in Cavazzani et al. (\cite{cava15}), provides the nightly correspondence between the satellite analysis and the SQM data analysis, and it is given by the complementary corellation coefficient in column 6. The annual average is $\epsilon^{Nightly}_{G-SQM}=2.8\%$ (e.g. If we associate an error of 1 night relative to 1 month, this means that an error of 1 night corresponds to about 3 per cent).\\
The same analysis is carried out for Asiago observing Table \ref{t3}. We calculate the monthly statistical uncertainty through the absolute value of the difference between column 2 and column 3, and the punctual uncertainty with the complementary corellation coefficient in column 4. The annual averages are $\epsilon_{A-SQM}=0.5\%$ and $\epsilon^{Nightly}_{A-SQM}=5.4\%$ respectively at Ekar observatory in 2018.\\
Finally, we estimated the total uncertainty through the uncertainty propagation, considering also the discrepancies between our algorithm and the ground data with the presence of particular climatic conditions (e.g. high humidity). 
We considered the ground data of a complex climatic conditions month for each site, as described in Sections \ref{c} and \ref{d}.
The punctual uncertainty between ground data and SQM data is $\epsilon_{LaSilla}=3.7\%$ while at Asiago the punctual uncertainty is $\epsilon_{Asiago}=5.9\%$. The total nightly uncertainty can be estimated through the formulas:

	\[\epsilon_{Total}=\sqrt{(\epsilon^{Nightly}_{G-SQM})^{2}+(\epsilon_{LaSilla})^{2}}
\]

and:

\[\epsilon_{Total}=\sqrt{(\epsilon^{Nightly}_{A-SQM})^{2}+(\epsilon_{Asiago})^{2}}
\]

getting an uncertainty of about 5.0\% at La Silla and 8.0\% at Asiago. We will deepen in a future work the correlation between the ground data and our algorithm for further validations and improvements.

\section{Discussion and conclusion}
\label{e}

The analysis of the night-time cloud cover is still an open problem. Satellites are primarily designed for daily analysis and have some limitations during the night.
In this paper we describe a new algorithm for the nocturnal analysis of cloud cover making use of the SQM data. 
We verified the results through the correlation with polar and geostationary satellites data and a sample of ground data in the most significant periods, namely the most climatically variable months.\\
Tables \ref{t1} and \ref{t3} show the results of this correlation.
This allows the use of a single instrument for measuring two factors that are important for the astronomical observations: sky brightness and and average cloud cover at night.  
The installation of the SQM instrument associated with the described algorithm will provide the extraction of an objective and low-cost statistics of the night cloud cover.
This would implement all current short and long term forecasting models. It has also been shown that the presence of clouds has the opposite effect on sites affected by light pollution or unaffected.
We also showed how also the Moon has an opposite effect in the two analysed conditions. The cloud cover with the full Moon reduces the sky brightness from magnitude 15.0 to 16.0 and lowers the SD at La Silla.
With full Moon, the clouds induce smaller variations than with new Moon at an uncontaminated site. They also filter moonlight, so the sky becomes darker.
At Asiago the magnitude changes from 18.5 to 18.0, and the maximum half-dispersion rises.	
In addition to astronomical applications, this study also explains the ALAN effects in all conditions with its consequences on flora and fauna.\\ 
Our algorithm shows how the SQM could be used to simultaneously detect the night sky brightness and nocturnal cloud cover. The empirical calibration of the threshold is a function of the magnitude detected by VIIRS (see Table \ref{t}) and of the lunar cycles (see Figure \ref{ath}).
We observed the SQM readings changes during intervals of 9 minutes for La Silla and 10 minutes for Asiago to exclude the gradual variations due to the presence of the Milky Way or the Moon. These intervals are also in agreement with the typical astronomical observation times.\\
The choice of 6-hour and 2-hour intervals also classifies the photometric and spectroscopic nights (see Tables \ref{t2} and \ref{t3}). The high temporal resolution of the SQM allows a real-time observation of night clouds improving the quality and the calibration of astronomical observations.
The high correlation between the satellite and our algorithm, as described in Section \ref{error}, also extrapolates the seasonal trends of the two sites (see Figures \ref{lspn} and \ref{apn}).\\
In conclusion we described a new algorithm for studying sky brightness and cloud cover during the night using only the SQM tool. This simple and cheap tool was extremely sensitive for the night clouds detection and very useful for collecting big data archives. This procedure can be applied to the entire SQM network contributing to the development and improvement of the astronomical telescopes scheduling.

\subsection{ACKNOWLEDGMENTS}

This activity is supported by the INAF (Istituto Nazionale di Astrofisica) funds allocated to the Premiale ADONI MIUR.
Most of the GOES data used in this paper are from CLASS
(Comprehensive Large Array-data Stewardship System), which is
an electronic library of NOAA environmental data. This website
provides capabilities for finding and obtaining data, particularly
NOAA's Geostationary Operational Environmental Satellite data.
MODIS data were provided by the \textit{Giovanni - Interactive Visualization and Analysis} website.
We also refer to the 3D atmospheric reconstruction project at Prato Piazza (Italy).
We thank Astronomer Researcher Marina Orio for her careful reading and advice to increase the clarity of this paper.
Funding from Italian Ministry of Education,
University and Research (MIUR) through the {\it Dipartimenti di
eccellenza project Science of the Universe}.
Finally we thank the support of the University of Padua for this search (Research grant, type A, Rep. 138, Prot. 3022, 26/10/2018).

\label{lastpage}

\end{document}